# *R*Leave: an *in silico* cross-validation protocol for transcript differential expression analysis.

**Authors:** Matheus Costa e Silva[1], Norma Lucena-Silva[2], Juliana Doblas Massaro[1*], Eduardo Antônio Donadi[1*].

[*]**These authors contributed equally as senior authors.**

**Institutions:**

[1]Division of Clinical Immunology, Department of Medicine, Ribeirão Preto Medical School, University of São Paulo, 14048-900, Ribeirão Preto, SP, Brazil.

[2]Laboratory of Immunogenetics, Department of Immunology, Aggeu Magalhães Institute, 50740-465, Recife, PE, Brazil.

**Abstract**

**Background and Objective:** The massive parallel sequencing technology facilitates new discoveries in terms of transcript differential analysis; however, all the new findings must be validated, since the diversity of transcript expression may impair the identification of the most relevant ones.

**Methods:** The proposed *R*Leave algorithm (implemented in the R environment) utilizes a combination of conventional analysis (classic edgeR) together with other mathematical methods (Leave-one-out sample technique and Decision Trees validation) to identify more relevant candidates to be *in vitro* or *in silico* validated.

**Results:** The *R*Leave protocol was tested using miRNome expression analysis of two sample groups (diabetes mellitus and acute lymphoblastic leukemia), and both had their most important differentially expressed miRNA confirmed by RT-qPCR.

**Conclusion:** This protocol is applicable in RNA-SEQ research, highlighting the most relevant transcripts for *in silico* and/or *in vitro* validation.

**Abbreviations:**

MPS: Massive parallel sequencing

RT-qPCR: Quantitative real-time polymerase chain reaction

FDR: False discovery rate

DE: Differential expression

DEA: Differential expression analysis

CPMs: counts per million

AUC: Area under the curve

T-ALL: children exhibiting T-cell acute lymphoblastic leukemia

B-ALL: children exhibiting B-cell acute lymphoblastic leukemia

miRNA: MicroRNA

miRNome: All miRNAs expressed in a specific sample

mRNA: Messenger RNA

T1D: Type 1 diabetes

T2D: Type 2 diabetes

# 1. Introduction

Data obtained by massive parallel sequencing (MPS) have disclosed a great number of differentially expressed transcripts that may be associated with disease pathogenesis, disease markers or both (Lam et al., 2015). Although MPS is a robust method for unveiling differentially expressed transcripts, they need to be validated by quantitative real-time polymerase chain reaction (RT-qPCR) or by *in vitro* functional analyses, which are expensive and time-consuming. The search for differentially expressed transcripts exhibiting more relevance than others has been primarily focused on the stringency of statistical values, including *P*value, False Discovery Rate (FDR) and LogFoldChange (LogFC). Even considering these statistical parameters, a great diversity of transcript expression between the analyzed samples may be observed within the same group of comparison. Therefore, a method intended to hierarchically classify these transcripts may circumvent the problem of transcript expression diversity, highlighting the more relevant ones.

We propose the *R*Leave algorithm, implemented in the R language, for the sequencing analysis of differentially expressed transcripts, aiming to reclassify the transcripts and to highlight the more relevant candidates to be used for defining disease pathogenesis or biomarkers. The *R*Leave algorithm was developed using edgeR package, applying the Leave-one-out sample technique (Efron, 1979), followed by an *in silico* validation of the differentially expressed transcripts using Decision Trees (Salzberg, 1994).

# 2. *R*Leave Implementation

## 2.1. Algorithm's method

The process consists of two branches: i) the upper branch executes a differential expression (DE) analysis (DEA), using the complex design of the edgeR package (Robinson et al., 2009), including all samples (N=number of samples) analyzed in each comparison (Anders et al., 2013). The output of this process consists of a table describing the counts per million (CPMs) of each transcript together with the statistical values (for instance, $P$value$\leq$0.05, False Discovery Rate (FDR$\leq$0.1) and LogFC>1.0); and ii) the lower branch executes a sample combination based on Leave-one-out technique (Efron, 1979), in which N DEAs are performed encompassing N-1 samples, i.e., each execution excludes only one of the tested samples and compute the differentially expressed miRNAs until all the samples are sequentially removed and replaced in the analysis. After finishing the N executions, all the resulting tables from the N DEAs performed in the lower branch step are grouped together with the DEA table resulted from the upper branch step. This assay will allow the identification of the transcripts that are more prevalent when considering all analysis generated by the method. The resulting final table will present the CPM and statistical values obtained in the upper branch DEA (used as reference) and the most prevalent transcripts when considering all analyzes (upper and lower branches), classified using as the major variable the maximum prevalence score.

The maximum prevalence score is calculated considering the following equation: the sum of how many times a differentially expressed miRNA appeared throughout all analyses (upper and lower branch analyses generated from $R$Leave sampling method), divided by the total number of analyses.

To validate the results, the final data of DEA performed by $R$Leave protocol is analyzed by a machine learning algorithm using the Decision Tree C4.5 (Salzberg, 1994), implemented as J48 in Weka 3.1.9 (Hall et al., 2009). The results are classified defining 10-fold cross-validation and the area under the curve (AUC) metrics. Several algorithms may be used to *in silico* validate DE transcripts; the choice for the final validation procedure was focused on the Decisions tree algorithm, which is based on low computational requirements (Salzberg, 1994).

The complete workflow of *R*Leave protocol is illustrated in **Figure 1**.

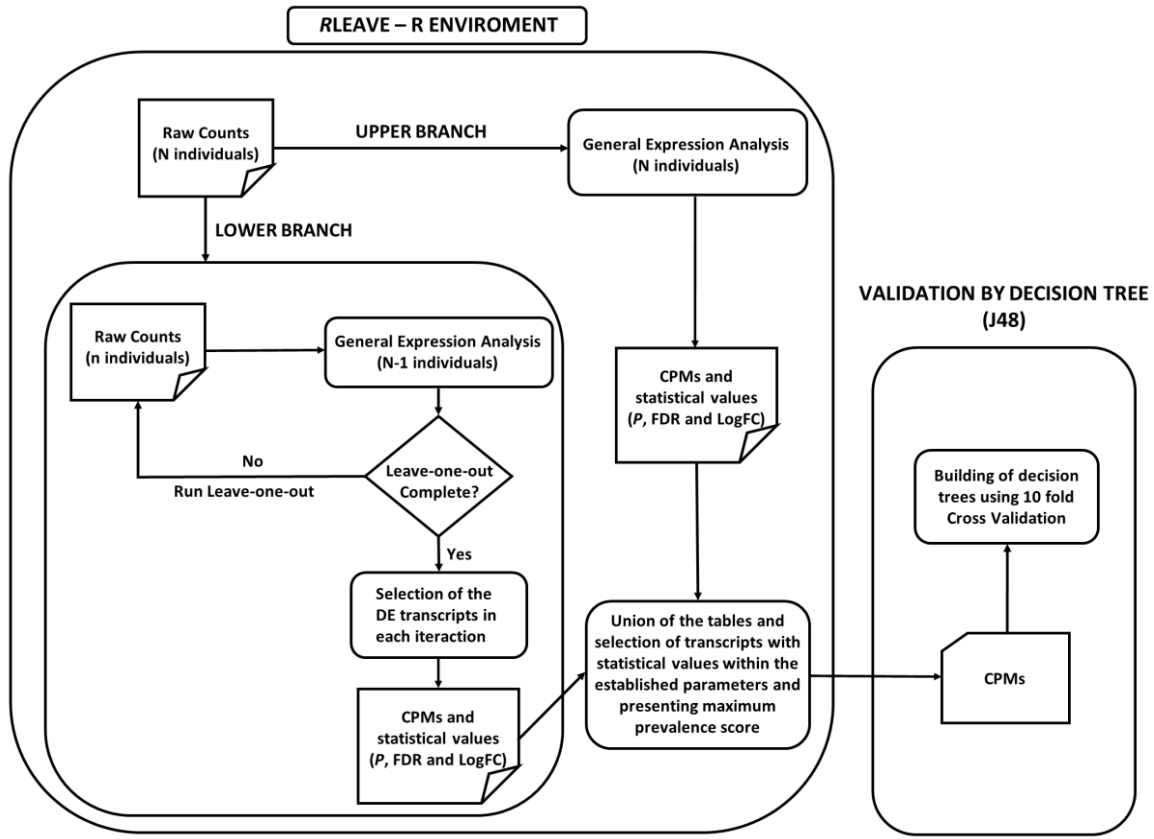

**Figure 1.** Workflow of the *R*Leave protocol. The upper branch performs the usual differential expression analysis, using the complex design of the edgeR package, showing the counts per million (CPM) and calculating the *P*value, the false dicovery rate (FDR) when applied and the logFoldChange (LogFC) (Anders et al., 2013). The lower branch performs the Leave-one-out procedure (Efron, 1979), considering the same statistical variables. After grouping the tables generated by the upper and lower branches, the Decision Tree analysis performs the validation of the most relevant transcripts (Middendorf et al., 2004).

## *2.2. Validation of the R*Leave *algorithm*

The feasibility of the *R*Leave algorithm was tested using diseases exhibiting distinct pathogenesis, acute lymphoblastic leukemias (Almeida et al., 2018) and diabetes mellitus (Massaro et al., 2019). Firstly, we used the raw data of miRNoma of children exhibiting T-cell acute lymphoblastic leukemia (T-ALL) *versus* children exhibiting B-cell acute lymphoblastic leukemia (B-ALL). The *R*Leave protocol revealed 16 DE miRNAs, and the validation by Random Forest classified the miR-29c-5p as the most relevant candidate to differentiate T-ALL from B-ALL. Finally, the same miRNA was also validated by RT-qPCR (Almeida et al., 2018). Secondly, we analyzed the raw data of miRNome of type 1 (T1D) and type 2 (T2D) diabetes mellitus patients presenting or not complications and performed several comparisons. For instance, in the comparison between T1D presenting retinopathy *versus* control; the *R*Leave disclosed three miRNAs, of which, the miR-143-3p was validated by RT-qPCR (Massaro et al., 2019). In both studies, the use of the *R*Leave algorithm permitted: i) an *in-silico* validation of DE transcripts, classifying the degree of importance of the variables in the ability to sample to differentiation, and ii) subsequent choice of better candidates for *in-vitro* validation.

## *2.3. Accessing the R*Leave *algorithm*

The protocol of the *R*Leave algorithm was submitted to Zenodo repository (freely downloaded at zenodo.org/record/3365736, registered at DOI 10.5281/zenodo.3365736).

To initialize the script of the algorithm ("Initializing Variables section): i) indicate the name of the table file, containing the sample names and identifiers; ii) insert the name of the raw count file, and indicate the minimum count number threshold; and iii) insert the differential expression variables (*P*value, FDR, and logFC). All variables may be modified according to your data and preferences, for instance, the user may or not consider FDR values for the selection of the relevant transcripts. After installing EdgeR, corrplot, and ggplot2 packages, simply set working directory in a folder containing the count files and sample identification and fill in the variable initialization fields set to run the script. The script will print every iteration from the protocol on the console, letting the user to know in which part of the execution the protocol is. More details can be found in the Zenodo repository (introductory text, counts.rda and groups.txt model files and commented source code).

## 3. Conclusion

The *R*Leave algorithm is a reliable resource for biomedical research, maintaining the same ability to differentiate samples considering the usual algorithms, and highlighting the most relevant transcripts to be validated *in silico* and *in vitro*.

All improvements on DEA can decrease costs of bench approaches and speed up research time on the discovery of biomarkers, allowing to the researcher the possibility to choose the stringency of the statistical values.

## 4. Funding

This study was supported by the "Conselho Nacional de Desenvolvimento Científico e Tecnológico" (CNPq, Brazil, Processes: #370246/2016-0 and #304931/2014-1) and "Coordenação de Aperfeiçoamento de Pessoal de Nível Superior" (CAPES, Brazil, Process: PROCAD #88881-068436/2014-1, Finance code #001).

## 5. Conflict of interest statement

The authors declare that there are no competing interests associated with this manuscript.